\documentclass[%
preprint,
amsmath,amssymb,
aps,
]{revtex4-1}

\usepackage{caption}
  
\usepackage{subcaption}
\usepackage{graphicx}
\usepackage{braket}
\usepackage{tikz}
\usetikzlibrary{decorations.pathmorphing}
\usetikzlibrary{shadows}

\usepackage{xcolor}

\begin{document}

\title{Plasmon-induced hot carriers from interband and intraband transitions in large noble metal nanoparticles}

\author{Hanwen Jin}
\affiliation{Department of Materials and the Thomas Young Centre for Theory and Simulation of Materials, Imperial College London, South Kensington Campus, London SW7 2AZ, UK\\}

\author{Juhan Matthias Kahk}
\affiliation{Institute of Physics, University of Tartu, W. Ostwaldi 1, 50411 Tartu, Estonia.\\}

\author{Dimitrios A. Papaconstantopoulos}
\affiliation{Department of Computational and Data Sciences
George Mason University, Fairfax, VA 22030, USA\\}

\author{Aires Ferreira}
\affiliation{Department of Physics and York Centre for Quantum Technologies, University of York, YO105DD, York, UK\\}

\author{Johannes Lischner}
\affiliation{Department of Materials and the Thomas Young Centre for Theory and Simulation of Materials, Imperial College London, South Kensington Campus, London SW7 2AZ, UK\\}

\begin{abstract}
Hot electrons generated from the decay of localized surface plasmons in metallic nanostructures have the potential to transform photocatalysis, photodetection and other optoelectronic applications. However, the understanding of hot-carrier generation in realistic nanostructures, in particular the relative importance of interband and intraband transitions, remains incomplete. Here we report theoretical predictions of hot-carrier generation rates in spherical nanoparticles of the noble metals silver, gold and copper with diameters up to 30 nanometers obtained from a novel atomistic linear-scaling approach. As the nanoparticle size increases the relative importance of interband transitions from d-bands to sp-bands relative to surface-enabled sp-band to sp-band transitions increases. We find that the hot-hole generation rate is characterized by a peak at the onset of the d-bands, while the position of the corresponding peak in the hot-electron distribution can be controlled through the illumination frequency. In contrast, intraband transitions give rise to hot electrons, but relatively cold holes. Importantly, increasing the dielectric constant of the environment removes hot carriers generated from interband transitions, while increasing the number of hot carriers from intraband transitions. The insights resulting from our work enable the design of nanoparticles for specific hot-carrier applications through their material composition, size and dielectric environment.
\end{abstract}

\maketitle

\section{Introduction}
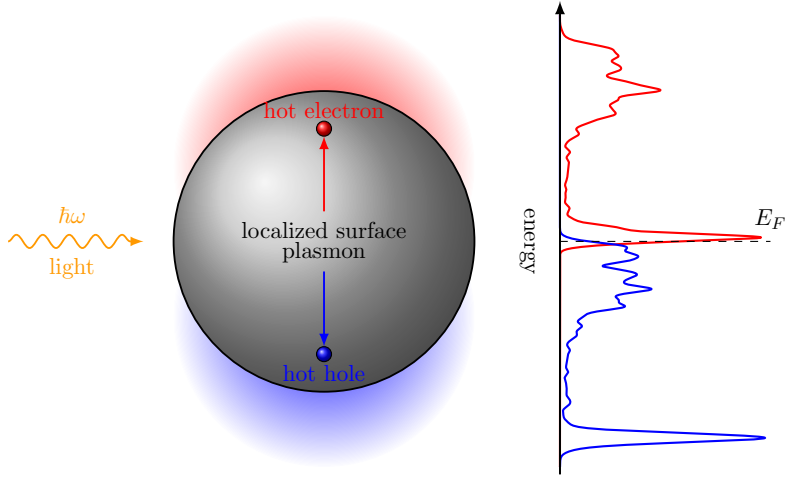
\begin{figure}
    \centering
    \begin{tikzpicture}

\definecolor{silver}{rgb}{0.75, 0.75, 0.75}
\definecolor{gold}{rgb}{1,.6,0}

\draw[-latex,gold,decorate, decoration=snake,line width=.7](-4.2,0) to (-2.4,0);
\node[gold,scale=.8] at (-3.35,-.35) {light};
\node[gold,scale=.8] at (-3.35,.35) {$\hbar\omega$};

\shade [inner color=red,outer color=white] (0,1) circle ( 2);
\shade [inner color=blue,outer color=white] (0,-1) circle ( 2);

\shadedraw [ball color= silver,line width=.7] (0,0) circle ( 2);
\shadedraw [ball color=red] (0,1.5) circle ( .1);
\shadedraw [ball color=blue] (0,-1.5) circle ( .1);
\node[red,scale=.8] at (0,1.75){hot electron};
\node[blue,scale=.8] at (0,-1.75){hot hole};
\draw[-latex,red,line width=.7] (0,.4)--(0,1.4);
\draw[-latex,blue,line width=.7] (0,-.4)--(0,-1.4);

\node[scale=.8] at (-0,.15) {localized surface};
\node[scale=.8] at (-0,-.15) {plasmon};

\node at (4.5,0) {\includegraphics{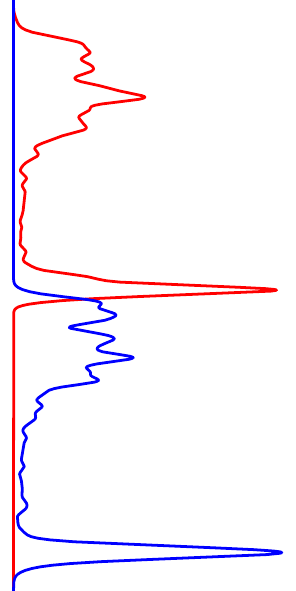}};
\draw[-latex,line width=.75](3.137,-3.1)--  (3.137,3.2);
\draw[dashed](3.135,0)  --(5.935,0);
\node[scale=.8] at (5.935,.3){$E_F$};
\node[scale=.8,rotate=-90] at (2.7,0){energy};

\end{tikzpicture}
    \caption{Schematic illustration of generation of a hot electron-hole pair resulting from the decay of a localized surface plasmon in a spherical metallic nanoparticle (left) and typical hot-carrier generation rates as function of energy (right). The Fermi level is denoted by $E_F$.}
    \label{fig:process}
\end{figure}

Metallic nanoparticles are highly efficient light absorbers when the light frequency matches the frequency of the localized surface plasmon (LSP)~\cite{Clavero2014,Aslam}. The LSP is a collective oscillation of the conduction electrons and gives rise to a large time-dependent dipole moment which couples strongly to the electric field of the light wave~\cite{maier_2007}. Importantly, the LSP is a strongly damped oscillation that decays on femtosecond time scales~\cite{Brown2015,Link1999}. Several decay channels are possible: a radiative decay resulting in the emission of light, a decay into one or more electron-hole pairs or decay mechanisms involving phonons~\cite{Brongersma2015}. The decay into one electron-hole pair is known as Landau damping, see Fig.~\ref{fig:process}, and is considered to be the most relevant for the generation of energetic or "hot" carriers at the surface of the nanoparticle~\cite{Govorov,Hartland}. Such hot carriers can be harnessed to drive chemical reactions~\cite{FUJISHIMA1972,Enrichi,Salvador2012,Yan2016,Thomann2011} or for photodetection~\cite{goykhman2011locally,li2017harvesting,Chalabi2014,Tang2020,Sun2019,Zhai2019,Zhu2021}, for example.

Despite the promise of hot-carrier devices for novel energy conversion and optoelectronic devices, many fundamental questions about the behaviour of hot carriers remain unsolved. For example, the question whether plasmonic photocatalysis is primarily facilitated by hot-carrier induced heating of the nanoparticle or hot-carrier transfer to the reactant has been the subject of much debate~\cite{Dubi2020,Sivan2019,Khurgin2019, DuChene2018,baffou2020applications}. Addressing these questions is crucial for the development and optimization of hot-carrier devices. To understand and identify the key driving force of plasmonic photocatalysis, a detailed microscopic understanding of hot-carrier processes in realistic nanostructures is required. This is very challenging to achieve, however, because quantum-mechanical first-principles approaches - while being highly accurate - can currently only be applied to periodic systems \cite{Sundararaman2014,Bernardi2015,Brown2015,Brown2016,zhang2019coexistence} or small nanoparticles \cite{RomnCastellanos2019,Rossi2020}. For example, Rossi and coworkers used ab initio time-dependent density functional theory to study hot-carrier generation in silver nanoparticles consisting of up to 561 atoms \cite{Rossi2020}. However, the radius of such a nanoparticle is only a few nanometers; much smaller than typical nanoparticles that are used in devices. To model larger nanoparticles, simplified electronic structure methods, such as the jellium \cite{Prodan2002,Manjavacas2014} or spherical-well approaches \cite{DalForno2018,Ranno2018,Manjavacas2014}, have been widely adopted. However, these methods only capture intraband transitions and lack a description of d-bands. As a consequence, they cannot describe the evolution of interband and intraband contributions to hot carrier generation as function of nanoparticle size, nor do they shed light on the role played by d-electrons in photocatalysis~\cite{Wilson2019,Tagliabue2020}.

In this paper, we use novel atomistic electronic structure techniques based on an accurate tight-binding Hamiltonian which includes d-states and reproduces the band structure of ab initio calculations~\cite{Joo2020,Joo2019,Ferreira2015,Weibe,Papaconstantopoulos2015} to study hot-carrier generation rates in large nanoparticles of silver, gold and copper with diameters up to 30 nm. We find that hot-carrier generation rates in small nanoparticles with diameters of a few nanometers exhibit a molecule-like behaviour with discrete peaks which evolve into a continuous distribution for larger nanoparticles. The hot-hole generation rates are characterized by a large peak at the onset of the d-bands stemming from interband d-to-sp band transitions \cite{RomnCastellanos2020}. The corresponding peak in the hot-electron generation rate is closer to the Fermi level, but its position can be controlled through the light frequency. Moreover, intraband transitions are shown to yield a second peak in the hot-hole generation rate which lies close to the Fermi level and a corresponding peak in the hot-electron rate at high energies. As the nanoparticle size increases, the interband transitions dominate increasingly over surface-enabled intraband transitions. Embedding the nanoparticle in an environment with a sufficiently large dielectric constant removes the contribution from interband hot carriers and enhances the one from intraband hot carriers. These insights from our work pave the way towards a mechanistic understanding of hot-carrier devices. For example, they enable experimentalists to design nanoparticles for specific oxidation and reduction reactions in photocatalysis.

\section{Results}

\subsection{Formalism}

When the nanoparticle is illuminated by light with frequency $\omega$, the rate of excited hot electrons with energy $E$ per unit volume can be obtained from Fermi's golden rule according to~\cite{Manjavacas2014,DalForno2018}
\begin{equation}\label{eq:Ne}
	N_e\left( E,\omega \right) =\frac{2}{V}\sum_{if}^{} \Gamma_{ if}\left( \omega \right) \delta\left( E-E_f;\sigma \right) ,
\end{equation} 
where $i$ and $f$ denote the initial and final states (with energies $E_i$ and $E_f$) of the light-induced electronic transition, $V$ is the nanoparticle volume and the factor of 2 accounts for spin degeneracy. Also, we defined the Gaussian broadened spectral function $\delta(x;\sigma)\equiv\frac{1}{2\pi\sigma^2}\exp\left(\frac{-x^2}{2\sigma^2}\right)$ which becomes a delta function in the limit of $\sigma\to0^+$. Here, $\sigma$ is a broadening parameter reflecting the finite quasiparticle linewidth. Finally, $\Gamma_{if} $ is given by 
\begin{equation}
	\Gamma_{ if}\left( \omega \right) =\frac{2 \pi}{\hbar}\left|\bra{f} \hat{\Phi}_{ tot} \left( \omega \right) \ket{i} \right|^2 \delta(E_f-E_i-\hbar\omega;\gamma)    f\left( E_{ i}  \right)\left( 1-f\left( E_f \right)  \right). 
\end{equation} 
Here, $f\left( E \right) $ denotes the Fermi-Dirac distribution function and $\hat{\Phi}_{tot}(\omega)$ is the total potential operator which consists of two contributions: the field arising from the external illumination and the induced field caused by the electronic response of the nanoparticle. Also, $\gamma$ denotes the typical linewidth of electronic transitions. In previous work~\cite{Manjavacas2014,DalForno2018} $\delta(E_f-E_i-\hbar\omega;\gamma)$ was often approximated by a Lorentzian function. However, we have found that this choice introduces artificial peaks near the Fermi level as a consequence of the slow decay of the Lorentzians. A similar expression for the rate of photo-excited hot holes can be obtained by swapping the indices of final and initial states in Eq.~\eqref{eq:Ne}.

To numerically evaluate the hot-carrier generation rates in nanoparticles, we employ the tight-binding approach to calculate the electronic states and their energies. Specifically, we assume that all nanoparticle wavefunctions can be expressed as linear combinations of the 3d, 4s and 4p atomic orbitals for Cu, 4d, 5s and 5p atomic orbitals for Ag and the 5d, 6s and 6p atomic orbitals for Au. The tight-binding Hamiltonian is based on an orthogonal two-center parametrization of ab initio density-functional theory calculations~\cite{Papaconstantopoulos2015}. The total potential is evaluated in the quasistatic approximation using experimentally measured bulk dielectric functions and the resulting matrix elements are calculated following the approach of Pedersen et al.~\cite{Pedersen2001}, see Methods for details.

The evaluation of Eq.~\eqref{eq:Ne} via exact diagonalization methods becomes impractical for large nanoparticles. To overcome this challenge, we exploit the observation that the hot-carrier generation rate, Eq.~\eqref{eq:Ne}, is a spectral quantity similar to the density of states $\rho(E)=\sum_n \delta(E - E_n)$. For such quantities, highly efficient numerical approaches have been recently developed which avoid the need to diagonalize large Hamiltonian matrices ~\cite{Joo2020,Joo2019,Ferreira2015,Weibe}. To harness these spectral methods, we write the hot-carrier rate as 
\begin{equation}\label{eq:spectral}
\begin{split}
    N_e(E,\omega) &= \frac{4\pi}{\hbar V} \int_{-\infty}^{\infty}d\mathcal{E}'\delta(E-\mathcal{E}';\sigma)\\ &\times\int_{-\infty}^{\infty}d\mathcal{E} \phi(\mathcal{E},\mathcal{E}',\omega) \delta(\mathcal{E}-\mathcal{E}' - \hbar\omega; \gamma) f(\mathcal{E}) (1-f(\mathcal{E}')),
\end{split}
\end{equation}
where $\phi(\mathcal{E},\mathcal{E}',\omega)=\sum_{if} |\langle f | \hat{\Phi}_{tot}(\omega)| i \rangle |^2 \delta(\mathcal{E} - E_i) \delta(\mathcal{E}'-E_f)$  can be conveniently expressed as the trace of the operator $\delta(\mathcal{E}-\hat{H}) \hat{\Phi}_{tot}(\omega) \delta(\mathcal{E}'-\hat{H}) \hat{\Phi}_{tot}(\omega)$. In order to apply the full machinery of spectral methods, we rescale and shift the energy variables $\mathcal{E}(\mathcal{E}')  \mapsto \varepsilon (\varepsilon^{\prime})$ and the  Hamiltonian $\hat H \mapsto \hat h$ so that the spectral weight is mapped into the interval $[-1:1]$, where first-kind Chebyshev polynomials, $T_n(\varepsilon)=\cos(n \arccos{\varepsilon})$ (with $n$ being a non-negative integer), form a complete set of orthogonal functions \cite{boyd01:CFS}. The spectral operator $\delta(\varepsilon-\hat h)$ can now be formally expressed as an infinite series of Chebyshev polynomials according to $\delta(\varepsilon-\hat h)=2/(\pi \sqrt {1 - \varepsilon^2}) \sum_{n=0}^{\infty} (1+\delta_{n0})^{-1}T_n(\hat h) T_n(\varepsilon)$. In practical calculations, this series is truncated after $N-1$ terms which induces unphysical Gibbs oscillations. These can be removed by multiplying each term in the series by with a coefficient of Jackson's kernel given by $J(n,N)=[(N-n)\cos(\pi n/N) + \sin(\pi n/N) \cot(\pi/N)]/N$~\cite{Jackson1911,Weibe} which effectively replaces each delta-function by a Gaussian (note that the width of these Gaussians is much smaller than the physical broadening parameters $\sigma$ and $\gamma$). Inserting these series into $\phi(\epsilon,\epsilon',\omega)$ gives
\begin{equation}
    \phi(\varepsilon,\varepsilon',\omega) \approx \frac{1}{E^2_-}\sum_{n=0}^{N-1}\sum_{m=0}^{N-1} \frac{4\mu_{mn}(\omega) T_m(\varepsilon) T_n(\varepsilon')}{\pi^2\sqrt{(1-\varepsilon^2)(1-\varepsilon'^2)}} \frac{J(n,N) J(m,N)}{(1+\delta_{n0})(1+\delta_{m0})},
\end{equation}
where we defined $\mu_{mn}(\omega)=\text{Tr} \left( T_m(\hat{h}) \hat{\Phi}_{tot}(\omega) T_n(\hat{h}) \hat{\Phi}_{tot}(\omega) \right)$ which can be computed efficiently exploiting the recurrence relation of Chebyshev polynomials and stochastic trace evaluation techniques ~\cite{Weibe}, see Methods for details.

\subsection{Hot-carrier generation rates}

\begin{figure}
    \centering
    \begin{tikzpicture}
    \node at (0,0) {\includegraphics{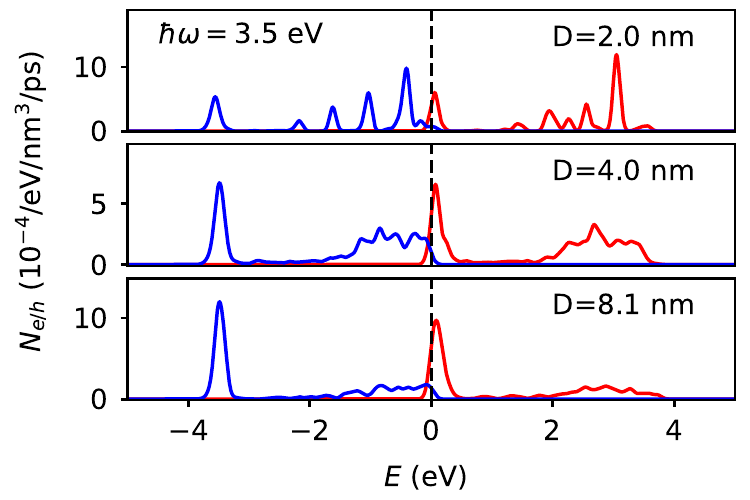}};
    \node at (3*2.54,0){\includegraphics{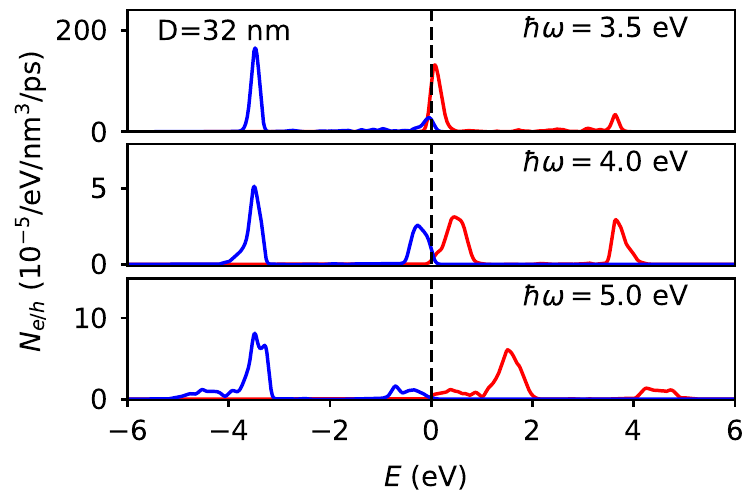}};
    \node at(-3.5,2.5) {\textbf{a}};
    \node at(-3.5+3*2.54,2.5) {\textbf{b}};
    \end{tikzpicture}
    \caption{Hot-carrier generation rates for spherical silver nanoparticles. a): Dependence of the hot-hole (blue) and hot-electron (red) rates on the nanoparticle diameter $D$ at the LSP frequency. b): Dependence of hot-carrier rates on the illumination frequency for a $D=32$~nm nanoparticle. The zero of energy is set to the Fermi level.}
    \label{fig:Ag}
\end{figure}

\begin{figure}
    \centering
    \begin{tikzpicture}
    \node at (0,0) {\includegraphics{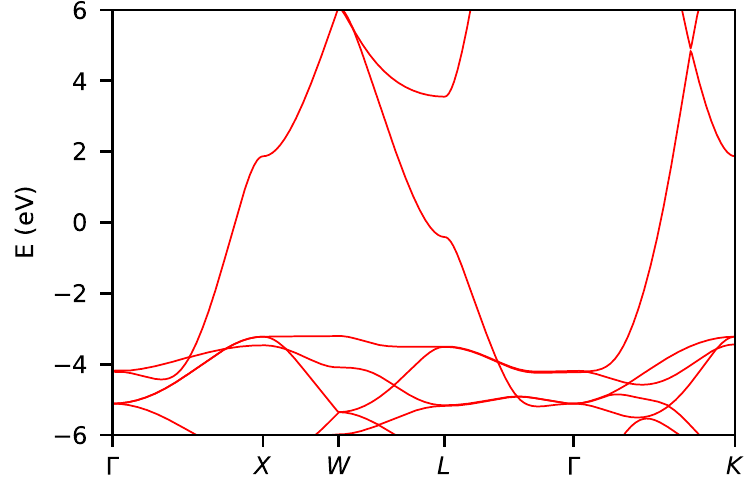}};
    \node at (3*2.54,0){\includegraphics{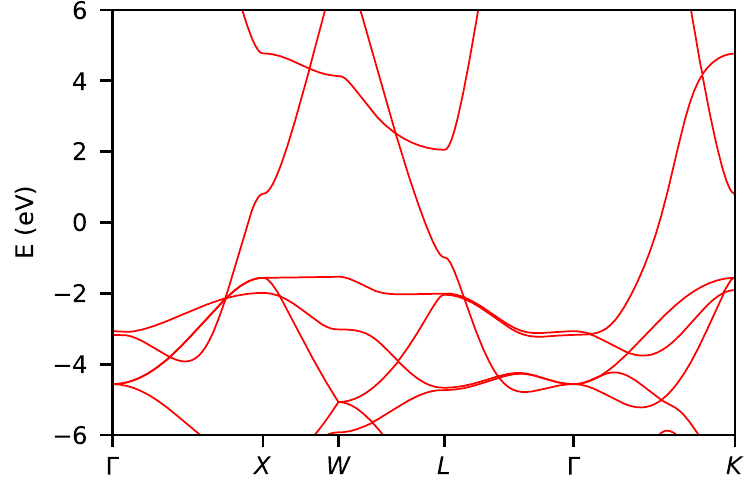}};
    \node at (0,-5.28) {\includegraphics{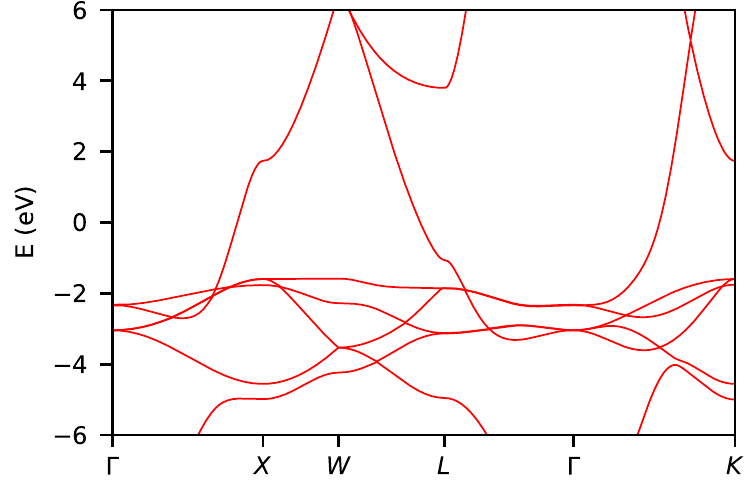}};
    \node at (3*2.54,-5.28){\includegraphics{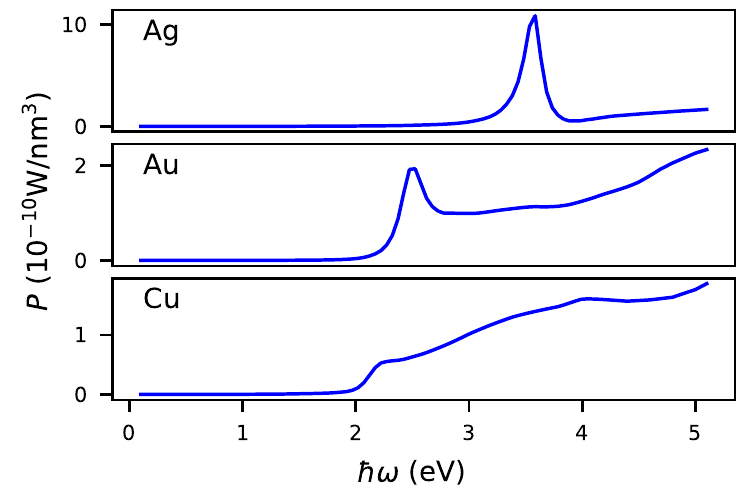}};
    \node at(-3.5,2.5) {\textbf{a}};
    \node at(-3.5+3*2.54,2.5) {\textbf{b}};
    \node at(-3.5,-2.5) {\textbf{c}};
    \node at(-3.5+3*2.54,-2.5) {\textbf{d}};
    \end{tikzpicture}
    \caption{Electronic band structures of (a) silver, (b) gold and (c) copper. The zero of energy is set to the Fermi level. (d) The power (for an illumination intensity of 1 mW/$\mu$m$^2$) absorbed by the spherical nanoparticles as function of photon energy.}
    \label{fig:band}
\end{figure}
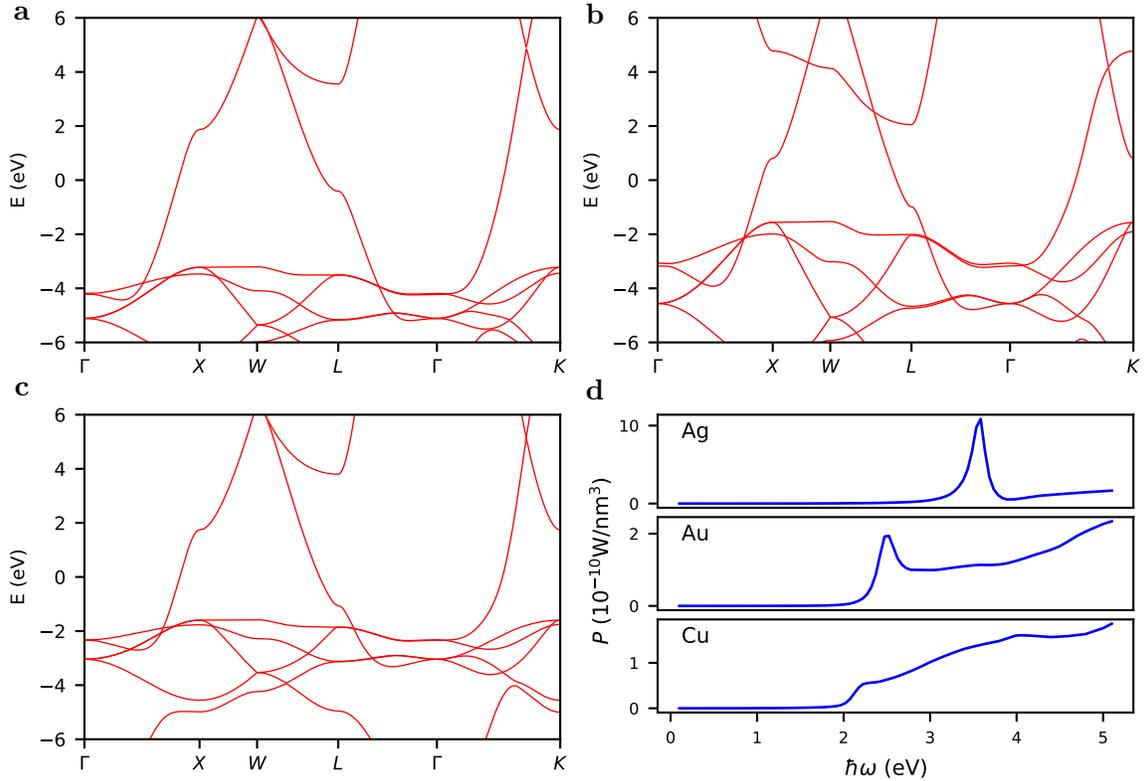

\begin{figure}
    \centering
    \begin{tikzpicture}
    \node at (0,0) {\includegraphics{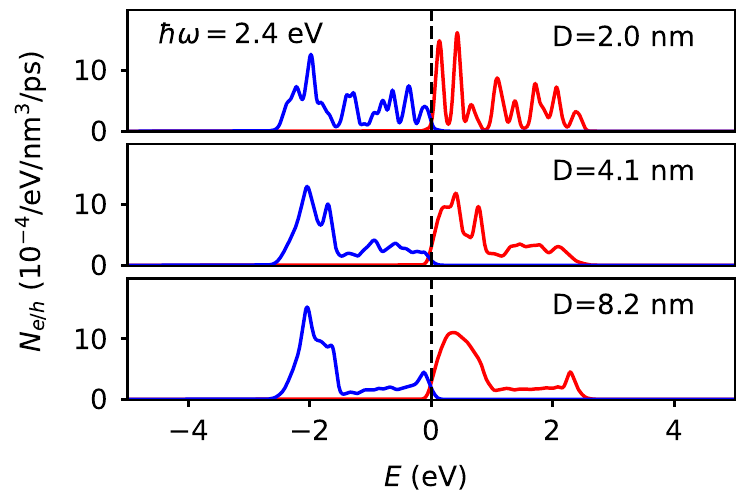}};
    \node at (3*2.54,0){\includegraphics{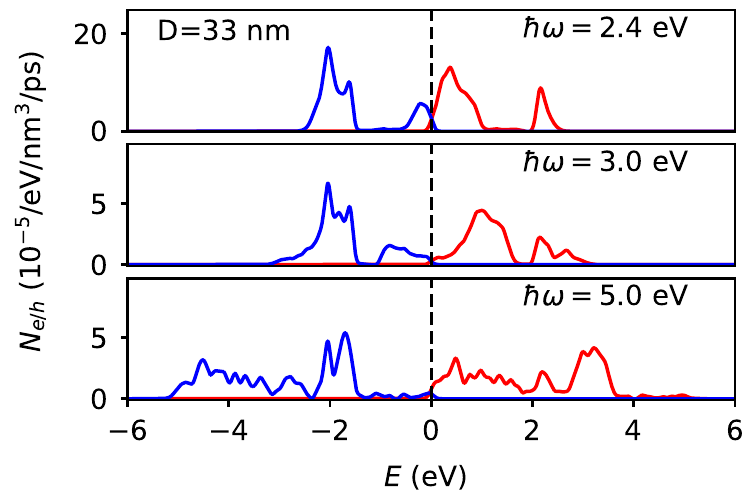}};
    \node at(-3.5,2.5) {\textbf{a}};
    \node at(-3.5+3*2.54,2.5) {\textbf{b}};
    \end{tikzpicture}
    \caption{Hot-carrier generation rates for spherical gold nanoparticles. (a): Dependence of the hot-hole (blue) and hot-electron (rate) rates on the nanoparticle diameter $D$ at the LSP frequency. (b): Dependence of hot-carrier rates on the illumination frequency for the $D=33$~nm nanoparticle. The zero of energy is set to the Fermi level.}
    \label{fig:Au}
\end{figure}

\begin{figure}
    \centering
    \begin{tikzpicture}
    \node at (0,0) {\includegraphics{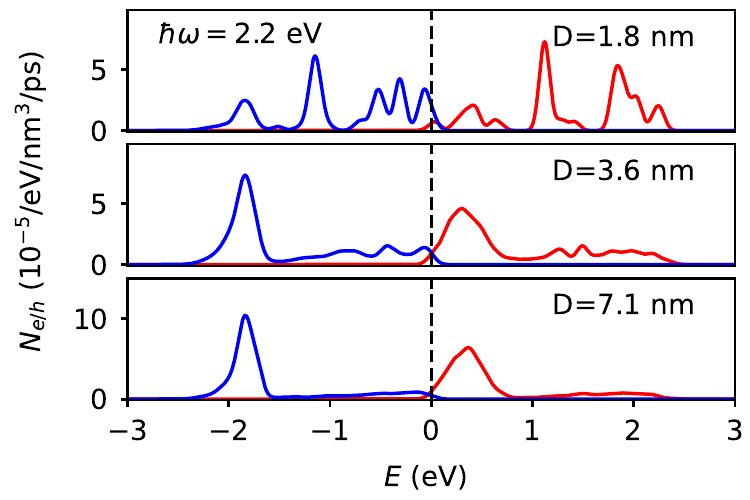}};
    \node at (3*2.54,0){\includegraphics{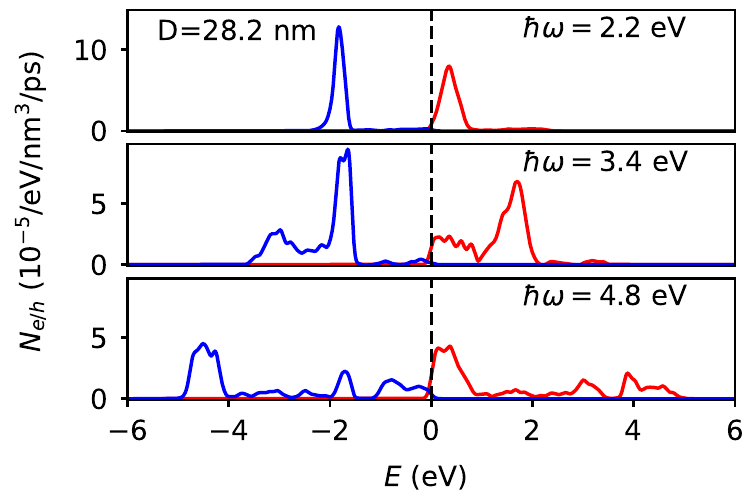}};
    \node at(-3.5,2.5) {\textbf{a}};
    \node at(-3.5+3*2.54,2.5) {\textbf{b}};
    \end{tikzpicture}
    \caption{Hot-carrier generation rates for spherical copper nanoparticles. (a): Dependence of the hot-hole (blue) and hot-electron (rate) rates on the nanoparticle diameter $D$ at the LSP frequency. (b): Dependence of hot-carrier rates on the illumination frequency for the $D=28.2$~nm nanoparticle. The zero of energy is set to the Fermi level.}
    \label{fig:Cu}
\end{figure}

We have calculated hot-carrier generation rates for spherical nanoparticles of Ag, Au and Cu containing up to 1,072,241 atoms (corresponding to diameters up to 32 nm (Ag), 33 nm (Au) and 28 nm (Cu)). 

Figure~\ref{fig:Ag}(a) shows the evolution of the hot-electron and hot-hole generation rates in Ag nanoparticles as function of the diameter when the nanoparticle is illuminated at the LSP frequency (3.5 eV in vacuum). For the smallest nanoparticle (D=2 nm), the hot-electron and hot-hole rates exhibit a series of discrete peaks characteristic of a molecule-like behaviour. At a diameter of 4 nm, we find that the hot-electron and hot-hole rates have evolved into smooth curves. The hot-hole rate has a sharp peak near -3.5 eV (relative to the Fermi level) and a second broader peak closer to the Fermi level. Because of energy conservation, the hot-electron rate also exhibits a sharp peak which is shifted from the sharp peak in the hot-hole rate by the LSP energy and lies close the Fermi level. The second peak in the hot-electron rate (which corresponds to the broad peak in the hot-hole rate near the Fermi level) is centered near 2.9 eV. The energy of the sharp peak in the hot-hole rate corresponds to the onset of the flat d-bands in Ag, see Fig.~\ref{fig:band}. When the nanoparticle is illuminated with light at the LSP frequency, interband transitions from the d-band to the sp-band which crosses the Fermi level can be induced. The broad peaks in the hot-electron and hot-hole rates, on the other hand, originate from transitions from sp-band states into other sp-band states. In contrast to the d-to-sp transitions, such transitions are forbidden in the bulk material and are only enabled by the presence of the surface. As the size of the nanoparticle increases, we expect the contribution of the surface-enabled transitions to decrease in comparison to the bulk transitions. Indeed, it can be seen that the size of the broad peak is significantly reduced for a diameter of 8.1 nm in comparison to the size of the sharp peak.

Next, we study the dependence of the hot-carrier generation rates on the light frequency. Fig.~\ref{fig:Ag}(b) shows the hot-electron and hot-hole rates of a Ag nanoparticle with a diameter of 32 nm (corresponding to 1,072,241 atoms) at illumination frequencies of 3.5 eV, 4.0 eV and 5.0 eV. Again, it can be seen that the hot-electron and the hot-hole rates exhibit two peaks. The strongest peak is in the hot-hole rate at around -3.5 eV (which, again, corresponds to the onset of the d-bands). The corresponding peak in the hot-electron rate is shifted by the illumination frequency, i.e. it moves to higher energies as the illumination frequency is increased. A second smaller peak in the hot-hole rate occurs just below the Fermi level where the density of occupied sp-band states is highest. This peak moves to lower energies as the illumination frequency is increased. The corresponding peak in the hot-electron rate is located near 4.0 eV and moves to slightly higher energies as the light frequency is increased. These results show that hot holes are predominantly produced in the d-band (as a result of d-to-sp transitions), but also near the Fermi level. In contrast, the energy of hot electrons generated by bulk transitions can be controlled through the illumination frequency, while the energy of surface-enabled hot electrons is very high. It can also be seen that the highest number of hot carriers are produced at the LSP frequency as a consequence of the large enhancement of the field intensity. At the LSP frequency, bulk transitions are much more frequent than surface-enabled transitions. In contrast, the relative contribution of surface-enabled hot carriers increases for higher illumination frequencies.

Figure~\ref{fig:Au} shows the hot-carrier rates for Au nanoparticles. At the LSP frequency (2.4 eV in vacuum), the hot-carrier rates again exhibit discrete peaks for small nanoparticles, which evolve into continuous distributions as the diameter increases, see Fig.~\ref{fig:Au}(a). The main peak in the hot-hole rate for the larger nanoparticles is located at -2.0 eV corresponding to the onset of d-bands in the material, see Fig.~\ref{fig:band}(b). A corresponding peak in the hot-electron rate is located just above the Fermi level. Both rates also exhibit a second smaller peak caused by surface-enabled sp-to-sp transitions. This second peak is just below the Fermi level level in the hot-hole rate and near +2.3 eV in the hot-electron rate. 

Considering next the evolution as function of light frequency for a nanoparticle of 33 nm diameter, see Fig.~\ref{fig:Au}(b), shows that the main peak in the hot-hole generation rate remains at -2.0 eV while the corresponding peak in the hot-electron rate moves to higher energies. In contrast to Ag, an additional hot-hole peak emerges at lower energies (near -4.5 eV for a light frequency of 5 eV) with a corresponding hot-electron peak just above the Fermi level. These peaks result from transitions from deeper lying d-bands to sp-bands. The hot-hole peak just below the Fermi level disappears as the light frequency increases.

After that, we consider Cu nanoparticles. In this system, the LSP peak at 2.2 eV is much less pronounced, see Figs.~\ref{fig:band}(c) and (d). As a consequence, the hot-carrier peaks at the LSP frequency are much lower than in Ag and Au, see Fig.~\ref{fig:Cu}(a). Similarly, to the case of Ag, we find a peak in the hot-hole rate at -1.8 eV corresponding to bulk d-to-sp transitions with a corresponding hot-electron peak just above the Fermi level. A second smaller peak in the hot hole rate arising from surface-enabled sp-to-sp transition is found just below the Fermi level and a corresponding hot electron peak near 2 eV. At higher frequencies, additional peaks at lower energies emerge in the hot-hole rate as a consequence of transitions from low-lying d-states. In particular, at a light frequency of 4.8 eV the strongest peak is located at -4.5 eV and a corresponding peak in the hot-electron rate is located at +0.4 eV. 

\begin{figure}
    \centering
    \includegraphics{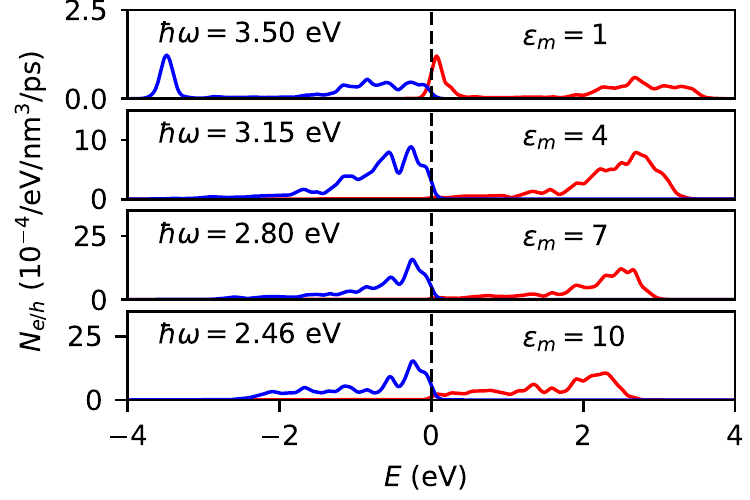}
    \caption{Hot-carrier distribution of a spherical silver nanoparticle with a diameter of 4 nm in various dielectric environments evaluated at the corresponding LSP frequencies. }
    \label{fig:Ag_em_dependence}
\end{figure}

Finally, we study the dependence of the hot-carrier generation rates on the dielectric constant of the surrounding medium. Fig.~\ref{fig:Ag_em_dependence} shows results for a silver nanoparticle with a diameter of 4 nm. As discussed above, in vacuum ($\epsilon_m = 1$), the hot-hole generation rate has a large peak at -3.5 eV due to interband transitions with a corresponding hot-electron peak near the Fermi level. For $\epsilon_m=4$, the LSP frequency has reduced to 3.15 eV. This is no longer sufficient to excite interband transitions and the hot-hole rate is characterized by a broad peak near the Fermi level, while the hot-hole rate exhibits a corresponding peak near 3 eV. Interestingly, the height of the intraband peak is increased compared to the case of vacuum. This increase is caused by an increase in the matrix elements~\cite{RomnCastellanos2020,roman2021dielectric} which are proportional to $1/\omega^2$~\cite{RomnCastellanos2020}. As the medium dielectric constant increases, the main peak of the hot-hole generation rate remains close to the Fermi level while the energy of the hot electrons is reduced.  

\section{Discussion}
We have reported atomistic quantum-mechanical calculations of plasmon-induced hot-carrier generation rates in large nanoparticles containing more than one million atoms. Accessing this size regime which is relevant for practical devices is possible through the use of highly efficient spectral methods based on an expansions of the hot-carrier generation rate in terms of Chebyshev polynomials. These advances allow us to study the evolution of the hot-carrier generation rate as function of nanoparticle diameter from a molecule-like regime characterized by discrete peaks to a continuous curve for diameters exceeding 4 nm. Moreover, the relative importance of bulk d-band to sp-band transitions compared to surface-enabled sp-band to sp-band transitions increases with increasing diameter. Interband transitions give rise to a large peak in the hot-hole generation rate located at the onset of d-band states. Such holes have energies of several electron volts relative to the Fermi level. The corresponding peak in the hot-electron rate lies closer to the Fermi level, but is tunable through the illumination frequency. In contrast, intraband transitions induce holes near the Fermi level and hot electrons with energies of several electron volts. Finally, the contribution to the hot-carrier distributions arising from interband transitions can be removed by increasing the dielectric constant of the surrounding medium which also enhances the matrix elements for intraband transitions. These insights are crucial for fabricating nanoparticles for inducing and optimizing specific chemical reactions. They also form the starting point for studying the thermalization of hot electrons which is the subject of future work.

\section{Methods}

Quasistatic approximation.--- We use the quasistatic approximation to evaluate the total electric potential experienced by electrons in the nanoparticle. If the electric field is parallel to the z-axis, the total potential inside the nanoparticle is given by 
\begin{equation}
    \Phi_{tot}(z,\omega) = -eE_0 \frac{3\epsilon_m}{\epsilon(\omega)+2\epsilon_m}z,
\end{equation}
where $E_0$ is the electric field strength of the external illumination, $\epsilon_m$ denotes the dielectric constant of the medium in which the nanoparticle is embedded and $\epsilon(\omega)$ denotes the bulk dielectric function of the material~\cite{Manjavacas2014,Otsuki2021,maier_2007}. Here, we use experimental bulk dielectric functions~\cite{haynes2015crc} to evaluate $\Phi_{tot}(z,\omega)$. In our calculations, we use $E_0=8.7\times 10^5$ V/m corresponding to an illumination intensity of 1 mW/$\mu$m$^2$. 

Matrix elements.--- To calculate the hot-carrier rates, we need to evaluate the matrix element of $\Phi_{tot}(z,\omega)$ between two nanoparticle states. Within the tight binding method, these states are expressed as linear combinations of atomic orbitals $|i,\alpha \rangle$, where the index $i$ labels the atom of the nanoparticle and the index $\alpha$ labels the orbitals on the atom. Following Pederson and workers~\cite{Pedersen2001}, the matrix element of the total potential between two atomic orbitals is evaluated as
\begin{equation}
    \langle j,\beta | \hat{\Phi}_{tot}(\omega) | i,\alpha \rangle = -eE_0 \frac{3\epsilon_m}{\epsilon(\omega)+2\epsilon_m} \left( z_j + d_{\alpha,\beta} \right) \delta_{ij},
\end{equation}
where $z_j$ denotes the z-coordinate of atom $j$ and $d_{\alpha,\beta}$ is the intra-atomic dipole matrix element between orbitals $|\alpha \rangle$ and $|\beta \rangle$. We calculate the intra-atomic matrix element using ab initio density-functional theory for isolated atoms as implemented in the electronic structure program FHI-aims~\cite{Blum2009}. In these calculations, the Perdew-Burke-Ernzerhof (PBE) exchange-correlation functional~\cite{Perdew1996} was used. Non-spin-polarized calculations were performed in all cases, resulting in a (n–1)d10 ns1 valence electron configuration for Cu, Ag, and Au. Scalar relativistic effects were included via the atomic ZORA formalism, and the FHI-aims default “tight” numerical basis sets were employed~\cite{vanLenthe1994,Blum2009}. In order to ensure that the calculated Kohn-Sham orbitals and the corresponding matrix elements match the canonical atomic orbital basis functions used in the tight binding calculations, a small external potential that lifts the degeneracies of the metal d- and p-orbitals was introduced in the DFT calculations. It was verified that the external potential did not affect the energies of the Kohn-Sham eigenstates by more than 0.01 eV. The non-vanishing matrix elements are shown in Table~\ref{tab:dipole}.

\begin{table}[]
    \centering
    \caption{Non-vanishing intra-atomic dipole matrix elements obtained from first-principles calculations, see text. }
    \label{tab:dipole}
    \begin{tabular}{ccccc}
         \hline
         $\alpha$&$\beta$&\multicolumn{3}{c}{$d_{\alpha,\beta}$ ($10^{-2}$ nm)}\\
         &&Ag&Au&Cu\\\hline
         $d_{zx}$&$p_x$&2.73&3.27 &2.31 \\
         $d_{yz}$&$p_z$&2.73&3.27 &2.31\\
         $d_{z^2}$&$p_z$&3.15&3.75&2.63\\
         $s$&$p_z$&9.36&8.95&8.55\\\hline
    \end{tabular}
\end{table}

Spectral approach.---To expand the spectral operator $\delta(E-\hat{H})$ in terms of Chebychev polynomials, the energy variable and the Hamiltonian must be rescaled and shifted such that their spectral weight lies in the interval $[-1;1]$. This is achieved by the following transformation: $\varepsilon=(\mathcal{E}-E_+)/E_-$ and $\hat{h}=(\hat{H}-E_{+})/E_{-}$ with $E_{\pm}=(E_{L} \pm E_{S})/2$ and $E_{L(S)}$ denotes the largest (smallest) energy level of the nanoparticle, which was approximated by the largest (smallest) energy of the bulk band structure. Truncating the series expansion of the spectral operator and multiplying each term with the coefficients of Jackson's kernel, effectively replaces the delta-function by a Gaussian with an energy-dependent width. The chosen value of $N=5,000$ gives rise to a maximum broadening of 12 meV for Ag, 9.5 meV for Au and 12 meV for Cu. These numerical broadenings are significantly smaller than the physical broadening parameters ($\sigma=50$~meV and $\gamma=60$~meV) that we use in our calculations.

Finally, evaluating $\phi(\epsilon,\epsilon',\omega)$ requires computation of the Chebyshev moments defined as $\mu_{mn}(\omega)=\textrm{Tr} \left( T_m(\hat{h}) \hat{\Phi}_{tot}(\omega) T_n(\hat{h}) \hat{\Phi}_{tot}(\omega) \right)$. This is achieved by using the recurrence relation of the Chebyshev polynomials and a stochastic trace evaluation technique which scales linearly with system size. In particular, we approximate
\begin{equation}
    \mu_{mn}(\omega) \approx \left| eE_0 \frac{3\epsilon_m}{\epsilon(\omega)+2\epsilon_m} \right|^2 \sum_{k=1}^K \langle k | T_m(\hat{h}) \hat{z} T_n(\hat{h}) \hat{z} | k \rangle , 
\end{equation}
where $\left\{|k \rangle \right\}$ denotes a set of random vectors. As the system size increases, the number $K$ of vectors in this set required for converged results decreases. In our calculations, we have used 100-10,000 random vectors depending on the diameter of nanoparticle which results in highly converged results.

\section{Acknowledgments}
JL and AF acknowledge funding from the Royal Society through a Royal Society University Research Fellowship. This project has received funding from the European Union's Horizon 2020 research and innovation programme under grant agreement No 892943. JMK acknowledges support from the Estonian Centre of Excellence in Research project “Advanced materials and high-technology devices for sustainable energetics, sensorics and nanoelectronics” TK141 (2014-2020.4.01.15-0011).


\begin{thebibliography}{52}%
\makeatletter
\providecommand \@ifxundefined [1]{%
 \@ifx{#1\undefined}
}%
\providecommand \@ifnum [1]{%
 \ifnum #1\expandafter \@firstoftwo
 \else \expandafter \@secondoftwo
 \fi
}%
\providecommand \@ifx [1]{%
 \ifx #1\expandafter \@firstoftwo
 \else \expandafter \@secondoftwo
 \fi
}%
\providecommand \natexlab [1]{#1}%
\providecommand \enquote  [1]{``#1''}%
\providecommand \bibnamefont  [1]{#1}%
\providecommand \bibfnamefont [1]{#1}%
\providecommand \citenamefont [1]{#1}%
\providecommand \href@noop [0]{\@secondoftwo}%
\providecommand \href [0]{\begingroup \@sanitize@url \@href}%
\providecommand \@href[1]{\@@startlink{#1}\@@href}%
\providecommand \@@href[1]{\endgroup#1\@@endlink}%
\providecommand \@sanitize@url [0]{\catcode `\\12\catcode `\$12\catcode
  `\&12\catcode `\#12\catcode `\^12\catcode `\_12\catcode `\%12\relax}%
\providecommand \@@startlink[1]{}%
\providecommand \@@endlink[0]{}%
\providecommand \url  [0]{\begingroup\@sanitize@url \@url }%
\providecommand \@url [1]{\endgroup\@href {#1}{\urlprefix }}%
\providecommand \urlprefix  [0]{URL }%
\providecommand \Eprint [0]{\href }%
\providecommand \doibase [0]{http://dx.doi.org/}%
\providecommand \selectlanguage [0]{\@gobble}%
\providecommand \bibinfo  [0]{\@secondoftwo}%
\providecommand \bibfield  [0]{\@secondoftwo}%
\providecommand \translation [1]{[#1]}%
\providecommand \BibitemOpen [0]{}%
\providecommand \bibitemStop [0]{}%
\providecommand \bibitemNoStop [0]{.\EOS\space}%
\providecommand \EOS [0]{\spacefactor3000\relax}%
\providecommand \BibitemShut  [1]{\csname bibitem#1\endcsname}%
\let\auto@bib@innerbib\@empty
\bibitem [{\citenamefont {Clavero}(2014)}]{Clavero2014}%
  \BibitemOpen
  \bibfield  {author} {\bibinfo {author} {\bibfnamefont {C.}~\bibnamefont
  {Clavero}},\ }\href@noop {} {\bibfield  {journal} {\bibinfo  {journal}
  {Nature Photonics}\ }\textbf {\bibinfo {volume} {8}},\ \bibinfo {pages} {95}
  (\bibinfo {year} {2014})}\BibitemShut {NoStop}%
\bibitem [{\citenamefont {Umar~Aslam}(2018)}]{Aslam}%
  \BibitemOpen
  \bibfield  {author} {\bibinfo {author} {\bibfnamefont {V.~G. R. e.~a.}\
  \bibnamefont {Umar~Aslam}},\ }\href {\doibase 10.1038/s41929-018-0138-x}
  {\bibfield  {journal} {\bibinfo  {journal} {Nature Catalysis}\ }\textbf
  {\bibinfo {volume} {1}},\ \bibinfo {pages} {656} (\bibinfo {year}
  {2018})}\BibitemShut {NoStop}%
\bibitem [{\citenamefont {Maier}(2007)}]{maier_2007}%
  \BibitemOpen
  \bibfield  {author} {\bibinfo {author} {\bibfnamefont {S.~A.}\ \bibnamefont
  {Maier}},\ }\href@noop {} {\emph {\bibinfo {title} {Plasmonics: Fundamentals
  and applications}}}\ (\bibinfo  {publisher} {Springer},\ \bibinfo {year}
  {2007})\BibitemShut {NoStop}%
\bibitem [{\citenamefont {Brown}\ \emph {et~al.}(2015)\citenamefont {Brown},
  \citenamefont {Sundararaman}, \citenamefont {Narang}, \citenamefont
  {Goddard},\ and\ \citenamefont {Atwater}}]{Brown2015}%
  \BibitemOpen
  \bibfield  {author} {\bibinfo {author} {\bibfnamefont {A.~M.}\ \bibnamefont
  {Brown}}, \bibinfo {author} {\bibfnamefont {R.}~\bibnamefont {Sundararaman}},
  \bibinfo {author} {\bibfnamefont {P.}~\bibnamefont {Narang}}, \bibinfo
  {author} {\bibfnamefont {W.~A.}\ \bibnamefont {Goddard}}, \ and\ \bibinfo
  {author} {\bibfnamefont {H.~A.}\ \bibnamefont {Atwater}},\ }\href@noop {}
  {\bibfield  {journal} {\bibinfo  {journal} {{ACS} Nano}\ }\textbf {\bibinfo
  {volume} {10}},\ \bibinfo {pages} {957} (\bibinfo {year} {2015})}\BibitemShut
  {NoStop}%
\bibitem [{\citenamefont {Link}\ and\ \citenamefont
  {El-Sayed}(1999)}]{Link1999}%
  \BibitemOpen
  \bibfield  {author} {\bibinfo {author} {\bibfnamefont {S.}~\bibnamefont
  {Link}}\ and\ \bibinfo {author} {\bibfnamefont {M.~A.}\ \bibnamefont
  {El-Sayed}},\ }\href@noop {} {\bibfield  {journal} {\bibinfo  {journal} {The
  Journal of Physical Chemistry B}\ }\textbf {\bibinfo {volume} {103}},\
  \bibinfo {pages} {8410} (\bibinfo {year} {1999})}\BibitemShut {NoStop}%
\bibitem [{\citenamefont {Brongersma}\ \emph {et~al.}(2015)\citenamefont
  {Brongersma}, \citenamefont {Halas},\ and\ \citenamefont
  {Nordlander}}]{Brongersma2015}%
  \BibitemOpen
  \bibfield  {author} {\bibinfo {author} {\bibfnamefont {M.~L.}\ \bibnamefont
  {Brongersma}}, \bibinfo {author} {\bibfnamefont {N.~J.}\ \bibnamefont
  {Halas}}, \ and\ \bibinfo {author} {\bibfnamefont {P.}~\bibnamefont
  {Nordlander}},\ }\href@noop {} {\bibfield  {journal} {\bibinfo  {journal}
  {Nature Nanotechnology}\ }\textbf {\bibinfo {volume} {10}},\ \bibinfo {pages}
  {25} (\bibinfo {year} {2015})}\BibitemShut {NoStop}%
\bibitem [{\citenamefont {Alexander O.~Govorov}(2006)}]{Govorov}%
  \BibitemOpen
  \bibfield  {author} {\bibinfo {author} {\bibfnamefont {W.~Z. e.~a.}\
  \bibnamefont {Alexander O.~Govorov}},\ }\href {\doibase
  10.1007/s11671-006-9015-7} {\bibfield  {journal} {\bibinfo  {journal}
  {Nanoscale Research Letters}\ }\textbf {\bibinfo {volume} {1}},\ \bibinfo
  {pages} {84} (\bibinfo {year} {2006})}\BibitemShut {NoStop}%
\bibitem [{\citenamefont {Gregory V.~Hartland}(2017)}]{Hartland}%
  \BibitemOpen
  \bibfield  {author} {\bibinfo {author} {\bibfnamefont {e.~a.}\ \bibnamefont
  {Gregory V.~Hartland}, \bibfnamefont {Lucas V.~Besteiro}},\ }\href {\doibase
  10.1021/acsenergylett.7b00333} {\bibfield  {journal} {\bibinfo  {journal}
  {ACS Energy Lett}\ }\textbf {\bibinfo {volume} {2}},\ \bibinfo {pages} {1641}
  (\bibinfo {year} {2017})}\BibitemShut {NoStop}%
\bibitem [{\citenamefont {Fujishima}\ and\ \citenamefont
  {Honda}(1972)}]{FUJISHIMA1972}%
  \BibitemOpen
  \bibfield  {author} {\bibinfo {author} {\bibfnamefont {A.}~\bibnamefont
  {Fujishima}}\ and\ \bibinfo {author} {\bibfnamefont {K.}~\bibnamefont
  {Honda}},\ }\href@noop {} {\bibfield  {journal} {\bibinfo  {journal}
  {Nature}\ }\textbf {\bibinfo {volume} {238}},\ \bibinfo {pages} {37}
  (\bibinfo {year} {1972})}\BibitemShut {NoStop}%
\bibitem [{\citenamefont {Enrichi}(2018)}]{Enrichi}%
  \BibitemOpen
  \bibfield  {author} {\bibinfo {author} {\bibfnamefont {F.}~\bibnamefont
  {Enrichi}},\ }\href {\doibase 10.1016/j.rser.2017.08.094} {\bibfield
  {journal} {\bibinfo  {journal} {Renewable and Sustainable Energy Reviews}\
  }\textbf {\bibinfo {volume} {82}},\ \bibinfo {pages} {1969} (\bibinfo {year}
  {2018})}\BibitemShut {NoStop}%
\bibitem [{\citenamefont {Salvador}\ \emph {et~al.}(2012)\citenamefont
  {Salvador}, \citenamefont {MacLeod}, \citenamefont {Hess}, \citenamefont
  {Kulkarni}, \citenamefont {Munechika}, \citenamefont {Chen},\ and\
  \citenamefont {Ginger}}]{Salvador2012}%
  \BibitemOpen
  \bibfield  {author} {\bibinfo {author} {\bibfnamefont {M.}~\bibnamefont
  {Salvador}}, \bibinfo {author} {\bibfnamefont {B.~A.}\ \bibnamefont
  {MacLeod}}, \bibinfo {author} {\bibfnamefont {A.}~\bibnamefont {Hess}},
  \bibinfo {author} {\bibfnamefont {A.~P.}\ \bibnamefont {Kulkarni}}, \bibinfo
  {author} {\bibfnamefont {K.}~\bibnamefont {Munechika}}, \bibinfo {author}
  {\bibfnamefont {J.~I.~L.}\ \bibnamefont {Chen}}, \ and\ \bibinfo {author}
  {\bibfnamefont {D.~S.}\ \bibnamefont {Ginger}},\ }\href@noop {} {\bibfield
  {journal} {\bibinfo  {journal} {{ACS} Nano}\ }\textbf {\bibinfo {volume}
  {6}},\ \bibinfo {pages} {10024} (\bibinfo {year} {2012})}\BibitemShut
  {NoStop}%
\bibitem [{\citenamefont {Yan}\ \emph {et~al.}(2016)\citenamefont {Yan},
  \citenamefont {Wang},\ and\ \citenamefont {Meng}}]{Yan2016}%
  \BibitemOpen
  \bibfield  {author} {\bibinfo {author} {\bibfnamefont {L.}~\bibnamefont
  {Yan}}, \bibinfo {author} {\bibfnamefont {F.}~\bibnamefont {Wang}}, \ and\
  \bibinfo {author} {\bibfnamefont {S.}~\bibnamefont {Meng}},\ }\href@noop {}
  {\bibfield  {journal} {\bibinfo  {journal} {{ACS} Nano}\ }\textbf {\bibinfo
  {volume} {10}},\ \bibinfo {pages} {5452} (\bibinfo {year}
  {2016})}\BibitemShut {NoStop}%
\bibitem [{\citenamefont {Thomann}\ \emph {et~al.}(2011)\citenamefont
  {Thomann}, \citenamefont {Pinaud}, \citenamefont {Chen}, \citenamefont
  {Clemens}, \citenamefont {Jaramillo},\ and\ \citenamefont
  {Brongersma}}]{Thomann2011}%
  \BibitemOpen
  \bibfield  {author} {\bibinfo {author} {\bibfnamefont {I.}~\bibnamefont
  {Thomann}}, \bibinfo {author} {\bibfnamefont {B.~A.}\ \bibnamefont {Pinaud}},
  \bibinfo {author} {\bibfnamefont {Z.}~\bibnamefont {Chen}}, \bibinfo {author}
  {\bibfnamefont {B.~M.}\ \bibnamefont {Clemens}}, \bibinfo {author}
  {\bibfnamefont {T.~F.}\ \bibnamefont {Jaramillo}}, \ and\ \bibinfo {author}
  {\bibfnamefont {M.~L.}\ \bibnamefont {Brongersma}},\ }\href@noop {}
  {\bibfield  {journal} {\bibinfo  {journal} {Nano Letters}\ }\textbf {\bibinfo
  {volume} {11}},\ \bibinfo {pages} {3440} (\bibinfo {year}
  {2011})}\BibitemShut {NoStop}%
\bibitem [{\citenamefont {Goykhman}\ \emph {et~al.}(2011)\citenamefont
  {Goykhman}, \citenamefont {Desiatov}, \citenamefont {Khurgin}, \citenamefont
  {Shappir},\ and\ \citenamefont {Levy}}]{goykhman2011locally}%
  \BibitemOpen
  \bibfield  {author} {\bibinfo {author} {\bibfnamefont {I.}~\bibnamefont
  {Goykhman}}, \bibinfo {author} {\bibfnamefont {B.}~\bibnamefont {Desiatov}},
  \bibinfo {author} {\bibfnamefont {J.}~\bibnamefont {Khurgin}}, \bibinfo
  {author} {\bibfnamefont {J.}~\bibnamefont {Shappir}}, \ and\ \bibinfo
  {author} {\bibfnamefont {U.}~\bibnamefont {Levy}},\ }\href@noop {} {\bibfield
   {journal} {\bibinfo  {journal} {Nano letters}\ }\textbf {\bibinfo {volume}
  {11}},\ \bibinfo {pages} {2219} (\bibinfo {year} {2011})}\BibitemShut
  {NoStop}%
\bibitem [{\citenamefont {Li}\ and\ \citenamefont
  {Valentine}(2017)}]{li2017harvesting}%
  \BibitemOpen
  \bibfield  {author} {\bibinfo {author} {\bibfnamefont {W.}~\bibnamefont
  {Li}}\ and\ \bibinfo {author} {\bibfnamefont {J.~G.}\ \bibnamefont
  {Valentine}},\ }\href@noop {} {\bibfield  {journal} {\bibinfo  {journal}
  {Nanophotonics}\ }\textbf {\bibinfo {volume} {6}},\ \bibinfo {pages} {177}
  (\bibinfo {year} {2017})}\BibitemShut {NoStop}%
\bibitem [{\citenamefont {Chalabi}\ \emph {et~al.}(2014)\citenamefont
  {Chalabi}, \citenamefont {Schoen},\ and\ \citenamefont
  {Brongersma}}]{Chalabi2014}%
  \BibitemOpen
  \bibfield  {author} {\bibinfo {author} {\bibfnamefont {H.}~\bibnamefont
  {Chalabi}}, \bibinfo {author} {\bibfnamefont {D.}~\bibnamefont {Schoen}}, \
  and\ \bibinfo {author} {\bibfnamefont {M.~L.}\ \bibnamefont {Brongersma}},\
  }\href@noop {} {\bibfield  {journal} {\bibinfo  {journal} {Nano Letters}\
  }\textbf {\bibinfo {volume} {14}},\ \bibinfo {pages} {1374} (\bibinfo {year}
  {2014})}\BibitemShut {NoStop}%
\bibitem [{\citenamefont {Tang}\ \emph {et~al.}(2020)\citenamefont {Tang},
  \citenamefont {Chen}, \citenamefont {Huang}, \citenamefont {Bright},
  \citenamefont {Meng}, \citenamefont {Liu},\ and\ \citenamefont
  {Wu}}]{Tang2020}%
  \BibitemOpen
  \bibfield  {author} {\bibinfo {author} {\bibfnamefont {H.}~\bibnamefont
  {Tang}}, \bibinfo {author} {\bibfnamefont {C.-J.}\ \bibnamefont {Chen}},
  \bibinfo {author} {\bibfnamefont {Z.}~\bibnamefont {Huang}}, \bibinfo
  {author} {\bibfnamefont {J.}~\bibnamefont {Bright}}, \bibinfo {author}
  {\bibfnamefont {G.}~\bibnamefont {Meng}}, \bibinfo {author} {\bibfnamefont
  {R.-S.}\ \bibnamefont {Liu}}, \ and\ \bibinfo {author} {\bibfnamefont
  {N.}~\bibnamefont {Wu}},\ }\href@noop {} {\bibfield  {journal} {\bibinfo
  {journal} {The Journal of Chemical Physics}\ }\textbf {\bibinfo {volume}
  {152}},\ \bibinfo {pages} {220901} (\bibinfo {year} {2020})}\BibitemShut
  {NoStop}%
\bibitem [{\citenamefont {Sun}\ \emph {et~al.}(2019)\citenamefont {Sun},
  \citenamefont {Zhang}, \citenamefont {Shao},\ and\ \citenamefont
  {Li}}]{Sun2019}%
  \BibitemOpen
  \bibfield  {author} {\bibinfo {author} {\bibfnamefont {Q.}~\bibnamefont
  {Sun}}, \bibinfo {author} {\bibfnamefont {C.}~\bibnamefont {Zhang}}, \bibinfo
  {author} {\bibfnamefont {W.}~\bibnamefont {Shao}}, \ and\ \bibinfo {author}
  {\bibfnamefont {X.}~\bibnamefont {Li}},\ }\href@noop {} {\bibfield  {journal}
  {\bibinfo  {journal} {{ACS} Omega}\ }\textbf {\bibinfo {volume} {4}},\
  \bibinfo {pages} {6020} (\bibinfo {year} {2019})}\BibitemShut {NoStop}%
\bibitem [{\citenamefont {Zhai}\ \emph {et~al.}(2019)\citenamefont {Zhai},
  \citenamefont {Chen}, \citenamefont {Ji}, \citenamefont {Ma}, \citenamefont
  {Wu}, \citenamefont {Li},\ and\ \citenamefont {Wang}}]{Zhai2019}%
  \BibitemOpen
  \bibfield  {author} {\bibinfo {author} {\bibfnamefont {Y.}~\bibnamefont
  {Zhai}}, \bibinfo {author} {\bibfnamefont {G.}~\bibnamefont {Chen}}, \bibinfo
  {author} {\bibfnamefont {J.}~\bibnamefont {Ji}}, \bibinfo {author}
  {\bibfnamefont {X.}~\bibnamefont {Ma}}, \bibinfo {author} {\bibfnamefont
  {Z.}~\bibnamefont {Wu}}, \bibinfo {author} {\bibfnamefont {Y.}~\bibnamefont
  {Li}}, \ and\ \bibinfo {author} {\bibfnamefont {Q.}~\bibnamefont {Wang}},\
  }\href@noop {} {\bibfield  {journal} {\bibinfo  {journal} {Nanotechnology}\
  }\textbf {\bibinfo {volume} {30}},\ \bibinfo {pages} {375201} (\bibinfo
  {year} {2019})}\BibitemShut {NoStop}%
\bibitem [{\citenamefont {Zhu}\ \emph {et~al.}(2021)\citenamefont {Zhu},
  \citenamefont {Xu}, \citenamefont {Yu},\ and\ \citenamefont
  {Wang}}]{Zhu2021}%
  \BibitemOpen
  \bibfield  {author} {\bibinfo {author} {\bibfnamefont {Y.}~\bibnamefont
  {Zhu}}, \bibinfo {author} {\bibfnamefont {H.}~\bibnamefont {Xu}}, \bibinfo
  {author} {\bibfnamefont {P.}~\bibnamefont {Yu}}, \ and\ \bibinfo {author}
  {\bibfnamefont {Z.}~\bibnamefont {Wang}},\ }\href {\doibase
  10.1063/5.0029050} {\bibfield  {journal} {\bibinfo  {journal} {Applied
  Physics Reviews}\ }\textbf {\bibinfo {volume} {8}},\ \bibinfo {pages}
  {021305} (\bibinfo {year} {2021})}\BibitemShut {NoStop}%
\bibitem [{\citenamefont {Dubi}\ \emph {et~al.}(2020)\citenamefont {Dubi},
  \citenamefont {Un},\ and\ \citenamefont {Sivan}}]{Dubi2020}%
  \BibitemOpen
  \bibfield  {author} {\bibinfo {author} {\bibfnamefont {Y.}~\bibnamefont
  {Dubi}}, \bibinfo {author} {\bibfnamefont {I.~W.}\ \bibnamefont {Un}}, \ and\
  \bibinfo {author} {\bibfnamefont {Y.}~\bibnamefont {Sivan}},\ }\href@noop {}
  {\bibfield  {journal} {\bibinfo  {journal} {Chemical Science}\ }\textbf
  {\bibinfo {volume} {11}},\ \bibinfo {pages} {5017} (\bibinfo {year}
  {2020})}\BibitemShut {NoStop}%
\bibitem [{\citenamefont {Sivan}\ \emph {et~al.}(2019)\citenamefont {Sivan},
  \citenamefont {Baraban}, \citenamefont {Un},\ and\ \citenamefont
  {Dubi}}]{Sivan2019}%
  \BibitemOpen
  \bibfield  {author} {\bibinfo {author} {\bibfnamefont {Y.}~\bibnamefont
  {Sivan}}, \bibinfo {author} {\bibfnamefont {J.}~\bibnamefont {Baraban}},
  \bibinfo {author} {\bibfnamefont {I.~W.}\ \bibnamefont {Un}}, \ and\ \bibinfo
  {author} {\bibfnamefont {Y.}~\bibnamefont {Dubi}},\ }\href@noop {} {\bibfield
   {journal} {\bibinfo  {journal} {Science}\ }\textbf {\bibinfo {volume} {364}}
  (\bibinfo {year} {2019})}\BibitemShut {NoStop}%
\bibitem [{\citenamefont {Khurgin}(2019)}]{Khurgin2019}%
  \BibitemOpen
  \bibfield  {author} {\bibinfo {author} {\bibfnamefont {J.~B.}\ \bibnamefont
  {Khurgin}},\ }\href@noop {} {\bibfield  {journal} {\bibinfo  {journal}
  {Faraday Discussions}\ }\textbf {\bibinfo {volume} {214}},\ \bibinfo {pages}
  {35} (\bibinfo {year} {2019})}\BibitemShut {NoStop}%
\bibitem [{\citenamefont {DuChene}\ \emph {et~al.}(2018)\citenamefont
  {DuChene}, \citenamefont {Tagliabue}, \citenamefont {Welch}, \citenamefont
  {Cheng},\ and\ \citenamefont {Atwater}}]{DuChene2018}%
  \BibitemOpen
  \bibfield  {author} {\bibinfo {author} {\bibfnamefont {J.~S.}\ \bibnamefont
  {DuChene}}, \bibinfo {author} {\bibfnamefont {G.}~\bibnamefont {Tagliabue}},
  \bibinfo {author} {\bibfnamefont {A.~J.}\ \bibnamefont {Welch}}, \bibinfo
  {author} {\bibfnamefont {W.-H.}\ \bibnamefont {Cheng}}, \ and\ \bibinfo
  {author} {\bibfnamefont {H.~A.}\ \bibnamefont {Atwater}},\ }\href@noop {}
  {\bibfield  {journal} {\bibinfo  {journal} {Nano Letters}\ }\textbf {\bibinfo
  {volume} {18}},\ \bibinfo {pages} {2545} (\bibinfo {year}
  {2018})}\BibitemShut {NoStop}%
\bibitem [{\citenamefont {Baffou}\ \emph {et~al.}(2020)\citenamefont {Baffou},
  \citenamefont {Cichos},\ and\ \citenamefont
  {Quidant}}]{baffou2020applications}%
  \BibitemOpen
  \bibfield  {author} {\bibinfo {author} {\bibfnamefont {G.}~\bibnamefont
  {Baffou}}, \bibinfo {author} {\bibfnamefont {F.}~\bibnamefont {Cichos}}, \
  and\ \bibinfo {author} {\bibfnamefont {R.}~\bibnamefont {Quidant}},\
  }\href@noop {} {\bibfield  {journal} {\bibinfo  {journal} {Nature Materials}\
  }\textbf {\bibinfo {volume} {19}},\ \bibinfo {pages} {946} (\bibinfo {year}
  {2020})}\BibitemShut {NoStop}%
\bibitem [{\citenamefont {Sundararaman}\ \emph {et~al.}(2014)\citenamefont
  {Sundararaman}, \citenamefont {Narang}, \citenamefont {Jermyn}, \citenamefont
  {III},\ and\ \citenamefont {Atwater}}]{Sundararaman2014}%
  \BibitemOpen
  \bibfield  {author} {\bibinfo {author} {\bibfnamefont {R.}~\bibnamefont
  {Sundararaman}}, \bibinfo {author} {\bibfnamefont {P.}~\bibnamefont
  {Narang}}, \bibinfo {author} {\bibfnamefont {A.~S.}\ \bibnamefont {Jermyn}},
  \bibinfo {author} {\bibfnamefont {W.~A.~G.}\ \bibnamefont {III}}, \ and\
  \bibinfo {author} {\bibfnamefont {H.~A.}\ \bibnamefont {Atwater}},\
  }\href@noop {} {\bibfield  {journal} {\bibinfo  {journal} {Nature
  Communications}\ }\textbf {\bibinfo {volume} {5}} (\bibinfo {year}
  {2014})}\BibitemShut {NoStop}%
\bibitem [{\citenamefont {Bernardi}\ \emph {et~al.}(2015)\citenamefont
  {Bernardi}, \citenamefont {Mustafa}, \citenamefont {Neaton},\ and\
  \citenamefont {Louie}}]{Bernardi2015}%
  \BibitemOpen
  \bibfield  {author} {\bibinfo {author} {\bibfnamefont {M.}~\bibnamefont
  {Bernardi}}, \bibinfo {author} {\bibfnamefont {J.}~\bibnamefont {Mustafa}},
  \bibinfo {author} {\bibfnamefont {J.~B.}\ \bibnamefont {Neaton}}, \ and\
  \bibinfo {author} {\bibfnamefont {S.~G.}\ \bibnamefont {Louie}},\ }\href@noop
  {} {\bibfield  {journal} {\bibinfo  {journal} {Nature Communications}\
  }\textbf {\bibinfo {volume} {6}} (\bibinfo {year} {2015})}\BibitemShut
  {NoStop}%
\bibitem [{\citenamefont {Brown}\ \emph {et~al.}(2016)\citenamefont {Brown},
  \citenamefont {Sundararaman}, \citenamefont {Narang}, \citenamefont
  {Goddard},\ and\ \citenamefont {Atwater}}]{Brown2016}%
  \BibitemOpen
  \bibfield  {author} {\bibinfo {author} {\bibfnamefont {A.~M.}\ \bibnamefont
  {Brown}}, \bibinfo {author} {\bibfnamefont {R.}~\bibnamefont {Sundararaman}},
  \bibinfo {author} {\bibfnamefont {P.}~\bibnamefont {Narang}}, \bibinfo
  {author} {\bibfnamefont {W.~A.}\ \bibnamefont {Goddard}}, \ and\ \bibinfo
  {author} {\bibfnamefont {H.~A.}\ \bibnamefont {Atwater}},\ }\href@noop {}
  {\bibfield  {journal} {\bibinfo  {journal} {Physical Review B}\ }\textbf
  {\bibinfo {volume} {94}} (\bibinfo {year} {2016})}\BibitemShut {NoStop}%
\bibitem [{\citenamefont {Zhang}\ \emph {et~al.}(2019)\citenamefont {Zhang},
  \citenamefont {Guan}, \citenamefont {Lischner}, \citenamefont {Meng},\ and\
  \citenamefont {Prezhdo}}]{zhang2019coexistence}%
  \BibitemOpen
  \bibfield  {author} {\bibinfo {author} {\bibfnamefont {J.}~\bibnamefont
  {Zhang}}, \bibinfo {author} {\bibfnamefont {M.}~\bibnamefont {Guan}},
  \bibinfo {author} {\bibfnamefont {J.}~\bibnamefont {Lischner}}, \bibinfo
  {author} {\bibfnamefont {S.}~\bibnamefont {Meng}}, \ and\ \bibinfo {author}
  {\bibfnamefont {O.~V.}\ \bibnamefont {Prezhdo}},\ }\href@noop {} {\bibfield
  {journal} {\bibinfo  {journal} {Nano letters}\ }\textbf {\bibinfo {volume}
  {19}},\ \bibinfo {pages} {3187} (\bibinfo {year} {2019})}\BibitemShut
  {NoStop}%
\bibitem [{\citenamefont {Castellanos}\ \emph {et~al.}(2019)\citenamefont
  {Castellanos}, \citenamefont {Hess},\ and\ \citenamefont
  {Lischner}}]{RomnCastellanos2019}%
  \BibitemOpen
  \bibfield  {author} {\bibinfo {author} {\bibfnamefont {L.~R.}\ \bibnamefont
  {Castellanos}}, \bibinfo {author} {\bibfnamefont {O.}~\bibnamefont {Hess}}, \
  and\ \bibinfo {author} {\bibfnamefont {J.}~\bibnamefont {Lischner}},\
  }\href@noop {} {\bibfield  {journal} {\bibinfo  {journal} {Communications
  Physics}\ }\textbf {\bibinfo {volume} {2}} (\bibinfo {year}
  {2019})}\BibitemShut {NoStop}%
\bibitem [{\citenamefont {Rossi}\ \emph {et~al.}(2020)\citenamefont {Rossi},
  \citenamefont {Erhart},\ and\ \citenamefont {Kuisma}}]{Rossi2020}%
  \BibitemOpen
  \bibfield  {author} {\bibinfo {author} {\bibfnamefont {T.~P.}\ \bibnamefont
  {Rossi}}, \bibinfo {author} {\bibfnamefont {P.}~\bibnamefont {Erhart}}, \
  and\ \bibinfo {author} {\bibfnamefont {M.}~\bibnamefont {Kuisma}},\
  }\href@noop {} {\bibfield  {journal} {\bibinfo  {journal} {{ACS} Nano}\
  }\textbf {\bibinfo {volume} {14}},\ \bibinfo {pages} {9963} (\bibinfo {year}
  {2020})}\BibitemShut {NoStop}%
\bibitem [{\citenamefont {Prodan}\ and\ \citenamefont
  {Nordlander}(2002)}]{Prodan2002}%
  \BibitemOpen
  \bibfield  {author} {\bibinfo {author} {\bibfnamefont {E.}~\bibnamefont
  {Prodan}}\ and\ \bibinfo {author} {\bibfnamefont {P.}~\bibnamefont
  {Nordlander}},\ }\href@noop {} {\bibfield  {journal} {\bibinfo  {journal}
  {Chemical Physics Letters}\ }\textbf {\bibinfo {volume} {352}},\ \bibinfo
  {pages} {140} (\bibinfo {year} {2002})}\BibitemShut {NoStop}%
\bibitem [{\citenamefont {Manjavacas}\ \emph {et~al.}(2014)\citenamefont
  {Manjavacas}, \citenamefont {Liu}, \citenamefont {Kulkarni},\ and\
  \citenamefont {Nordlander}}]{Manjavacas2014}%
  \BibitemOpen
  \bibfield  {author} {\bibinfo {author} {\bibfnamefont {A.}~\bibnamefont
  {Manjavacas}}, \bibinfo {author} {\bibfnamefont {J.~G.}\ \bibnamefont {Liu}},
  \bibinfo {author} {\bibfnamefont {V.}~\bibnamefont {Kulkarni}}, \ and\
  \bibinfo {author} {\bibfnamefont {P.}~\bibnamefont {Nordlander}},\
  }\href@noop {} {\bibfield  {journal} {\bibinfo  {journal} {{ACS} Nano}\
  }\textbf {\bibinfo {volume} {8}},\ \bibinfo {pages} {7630} (\bibinfo {year}
  {2014})}\BibitemShut {NoStop}%
\bibitem [{\citenamefont {Forno}\ \emph {et~al.}(2018)\citenamefont {Forno},
  \citenamefont {Ranno},\ and\ \citenamefont {Lischner}}]{DalForno2018}%
  \BibitemOpen
  \bibfield  {author} {\bibinfo {author} {\bibfnamefont {S.~D.}\ \bibnamefont
  {Forno}}, \bibinfo {author} {\bibfnamefont {L.}~\bibnamefont {Ranno}}, \ and\
  \bibinfo {author} {\bibfnamefont {J.}~\bibnamefont {Lischner}},\ }\href@noop
  {} {\bibfield  {journal} {\bibinfo  {journal} {The Journal of Physical
  Chemistry C}\ }\textbf {\bibinfo {volume} {122}},\ \bibinfo {pages} {8517}
  (\bibinfo {year} {2018})}\BibitemShut {NoStop}%
\bibitem [{\citenamefont {Ranno}\ \emph {et~al.}(2018)\citenamefont {Ranno},
  \citenamefont {Forno},\ and\ \citenamefont {Lischner}}]{Ranno2018}%
  \BibitemOpen
  \bibfield  {author} {\bibinfo {author} {\bibfnamefont {L.}~\bibnamefont
  {Ranno}}, \bibinfo {author} {\bibfnamefont {S.~D.}\ \bibnamefont {Forno}}, \
  and\ \bibinfo {author} {\bibfnamefont {J.}~\bibnamefont {Lischner}},\
  }\href@noop {} {\bibfield  {journal} {\bibinfo  {journal} {npj Computational
  Materials}\ }\textbf {\bibinfo {volume} {4}} (\bibinfo {year}
  {2018})}\BibitemShut {NoStop}%
\bibitem [{\citenamefont {Wilson}\ \emph {et~al.}(2019)\citenamefont {Wilson},
  \citenamefont {Mohan},\ and\ \citenamefont {Jain}}]{Wilson2019}%
  \BibitemOpen
  \bibfield  {author} {\bibinfo {author} {\bibfnamefont {A.~J.}\ \bibnamefont
  {Wilson}}, \bibinfo {author} {\bibfnamefont {V.}~\bibnamefont {Mohan}}, \
  and\ \bibinfo {author} {\bibfnamefont {P.~K.}\ \bibnamefont {Jain}},\
  }\href@noop {} {\bibfield  {journal} {\bibinfo  {journal} {The Journal of
  Physical Chemistry C}\ }\textbf {\bibinfo {volume} {123}},\ \bibinfo {pages}
  {29360} (\bibinfo {year} {2019})}\BibitemShut {NoStop}%
\bibitem [{\citenamefont {Tagliabue}\ \emph {et~al.}(2020)\citenamefont
  {Tagliabue}, \citenamefont {DuChene}, \citenamefont {Abdellah}, \citenamefont
  {Habib}, \citenamefont {Gosztola}, \citenamefont {Hattori}, \citenamefont
  {Cheng}, \citenamefont {Zheng}, \citenamefont {Canton}, \citenamefont
  {Sundararaman}, \citenamefont {S{\'{a}}},\ and\ \citenamefont
  {Atwater}}]{Tagliabue2020}%
  \BibitemOpen
  \bibfield  {author} {\bibinfo {author} {\bibfnamefont {G.}~\bibnamefont
  {Tagliabue}}, \bibinfo {author} {\bibfnamefont {J.~S.}\ \bibnamefont
  {DuChene}}, \bibinfo {author} {\bibfnamefont {M.}~\bibnamefont {Abdellah}},
  \bibinfo {author} {\bibfnamefont {A.}~\bibnamefont {Habib}}, \bibinfo
  {author} {\bibfnamefont {D.~J.}\ \bibnamefont {Gosztola}}, \bibinfo {author}
  {\bibfnamefont {Y.}~\bibnamefont {Hattori}}, \bibinfo {author} {\bibfnamefont
  {W.-H.}\ \bibnamefont {Cheng}}, \bibinfo {author} {\bibfnamefont
  {K.}~\bibnamefont {Zheng}}, \bibinfo {author} {\bibfnamefont {S.~E.}\
  \bibnamefont {Canton}}, \bibinfo {author} {\bibfnamefont {R.}~\bibnamefont
  {Sundararaman}}, \bibinfo {author} {\bibfnamefont {J.}~\bibnamefont
  {S{\'{a}}}}, \ and\ \bibinfo {author} {\bibfnamefont {H.~A.}\ \bibnamefont
  {Atwater}},\ }\href@noop {} {\bibfield  {journal} {\bibinfo  {journal}
  {Nature Materials}\ }\textbf {\bibinfo {volume} {19}},\ \bibinfo {pages}
  {1312} (\bibinfo {year} {2020})}\BibitemShut {NoStop}%
\bibitem [{\citenamefont {Jo$\tilde{\text{a}}$o}\ \emph
  {et~al.}(2020)\citenamefont {Jo$\tilde{\text{a}}$o}, \citenamefont
  {Andelkovic}, \citenamefont {Covaci}, \citenamefont {Rappoport},
  \citenamefont {Lopes},\ and\ \citenamefont {Ferreira}}]{Joo2020}%
  \BibitemOpen
  \bibfield  {author} {\bibinfo {author} {\bibfnamefont {S.~M.}\ \bibnamefont
  {Jo$\tilde{\text{a}}$o}}, \bibinfo {author} {\bibfnamefont {M.}~\bibnamefont
  {Andelkovic}}, \bibinfo {author} {\bibfnamefont {L.}~\bibnamefont {Covaci}},
  \bibinfo {author} {\bibfnamefont {T.~G.}\ \bibnamefont {Rappoport}}, \bibinfo
  {author} {\bibfnamefont {J.~M. V.~P.}\ \bibnamefont {Lopes}}, \ and\ \bibinfo
  {author} {\bibfnamefont {A.}~\bibnamefont {Ferreira}},\ }\href@noop {}
  {\bibfield  {journal} {\bibinfo  {journal} {Royal Society Open Science}\
  }\textbf {\bibinfo {volume} {7}},\ \bibinfo {pages} {191809} (\bibinfo {year}
  {2020})}\BibitemShut {NoStop}%
\bibitem [{\citenamefont {João}\ and\ \citenamefont {Lopes}(2019)}]{Joo2019}%
  \BibitemOpen
  \bibfield  {author} {\bibinfo {author} {\bibfnamefont {S.~M.}\ \bibnamefont
  {João}}\ and\ \bibinfo {author} {\bibfnamefont {J.~M. V.~P.}\ \bibnamefont
  {Lopes}},\ }\href@noop {} {\bibfield  {journal} {\bibinfo  {journal} {Journal
  of Physics: Condensed Matter}\ }\textbf {\bibinfo {volume} {32}},\ \bibinfo
  {pages} {125901} (\bibinfo {year} {2019})}\BibitemShut {NoStop}%
\bibitem [{\citenamefont {Ferreira}\ and\ \citenamefont
  {Mucciolo}(2015)}]{Ferreira2015}%
  \BibitemOpen
  \bibfield  {author} {\bibinfo {author} {\bibfnamefont {A.}~\bibnamefont
  {Ferreira}}\ and\ \bibinfo {author} {\bibfnamefont {E.~R.}\ \bibnamefont
  {Mucciolo}},\ }\href@noop {} {\bibfield  {journal} {\bibinfo  {journal}
  {Physical Review Letters}\ }\textbf {\bibinfo {volume} {115}} (\bibinfo
  {year} {2015})}\BibitemShut {NoStop}%
\bibitem [{\citenamefont {Weiße}\ \emph {et~al.}(2006)\citenamefont {Weiße},
  \citenamefont {Wellein}, \citenamefont {Alvermann},\ and\ \citenamefont
  {Fehske}}]{Weibe}%
  \BibitemOpen
  \bibfield  {author} {\bibinfo {author} {\bibfnamefont {A.}~\bibnamefont
  {Weiße}}, \bibinfo {author} {\bibfnamefont {G.}~\bibnamefont {Wellein}},
  \bibinfo {author} {\bibfnamefont {A.}~\bibnamefont {Alvermann}}, \ and\
  \bibinfo {author} {\bibfnamefont {H.}~\bibnamefont {Fehske}},\ }\href@noop {}
  {\ \textbf {\bibinfo {volume} {78}},\ \bibinfo {pages} {275} (\bibinfo {year}
  {2006})}\BibitemShut {NoStop}%
\bibitem [{\citenamefont
  {Papaconstantopoulos}(2015)}]{Papaconstantopoulos2015}%
  \BibitemOpen
  \bibfield  {author} {\bibinfo {author} {\bibfnamefont {D.~A.}\ \bibnamefont
  {Papaconstantopoulos}},\ }\href@noop {} {\emph {\bibinfo {title} {Handbook of
  the Band Structure of Elemental Solids}}}\ (\bibinfo  {publisher} {Springer
  {US}},\ \bibinfo {year} {2015})\BibitemShut {NoStop}%
\bibitem [{\citenamefont {Castellanos}\ \emph {et~al.}(2020)\citenamefont
  {Castellanos}, \citenamefont {Kahk}, \citenamefont {Hess},\ and\
  \citenamefont {Lischner}}]{RomnCastellanos2020}%
  \BibitemOpen
  \bibfield  {author} {\bibinfo {author} {\bibfnamefont {L.~R.}\ \bibnamefont
  {Castellanos}}, \bibinfo {author} {\bibfnamefont {J.~M.}\ \bibnamefont
  {Kahk}}, \bibinfo {author} {\bibfnamefont {O.}~\bibnamefont {Hess}}, \ and\
  \bibinfo {author} {\bibfnamefont {J.}~\bibnamefont {Lischner}},\ }\href@noop
  {} {\bibfield  {journal} {\bibinfo  {journal} {The Journal of Chemical
  Physics}\ }\textbf {\bibinfo {volume} {152}},\ \bibinfo {pages} {104111}
  (\bibinfo {year} {2020})}\BibitemShut {NoStop}%
\bibitem [{\citenamefont {Pedersen}\ \emph {et~al.}(2001)\citenamefont
  {Pedersen}, \citenamefont {Pedersen},\ and\ \citenamefont
  {Kriestensen}}]{Pedersen2001}%
  \BibitemOpen
  \bibfield  {author} {\bibinfo {author} {\bibfnamefont {T.~G.}\ \bibnamefont
  {Pedersen}}, \bibinfo {author} {\bibfnamefont {K.}~\bibnamefont {Pedersen}},
  \ and\ \bibinfo {author} {\bibfnamefont {T.~B.}\ \bibnamefont
  {Kriestensen}},\ }\href@noop {} {\bibfield  {journal} {\bibinfo  {journal}
  {Physical Review B}\ }\textbf {\bibinfo {volume} {63}} (\bibinfo {year}
  {2001})}\BibitemShut {NoStop}%
\bibitem [{\citenamefont {Boyd}(2001)}]{boyd01:CFS}%
  \BibitemOpen
  \bibfield  {author} {\bibinfo {author} {\bibfnamefont {J.~P.}\ \bibnamefont
  {Boyd}},\ }\href@noop {} {\emph {\bibinfo {title} {{Chebyshev} and {Fourier}
  Spectral Methods}}},\ \bibinfo {edition} {2nd}\ ed.\ (\bibinfo  {publisher}
  {Dover},\ \bibinfo {address} {Mineola, New York},\ \bibinfo {year}
  {2001})\BibitemShut {NoStop}%
\bibitem [{\citenamefont {Jackson}(1911)}]{Jackson1911}%
  \BibitemOpen
  \bibfield  {author} {\bibinfo {author} {\bibfnamefont {D.}~\bibnamefont
  {Jackson}},\ }\href@noop {} {\emph {\bibinfo {title} {\"Uber die Genauigkeit
  der Ann\"aherung stetiger Funktionen durch ganze rationale Funktionen
  gegebenen Grades und trigonometrische summen gegebener Ordnung}}}\ (\bibinfo
  {publisher} {Dieterich'schen Universit\"at -- Buchdruckerei},\ \bibinfo
  {year} {1911})\BibitemShut {NoStop}%
\bibitem [{\citenamefont {Rom\'an~Castellanos}\ \emph
  {et~al.}(2021)\citenamefont {Rom\'an~Castellanos}, \citenamefont {Hess},\
  and\ \citenamefont {Lischner}}]{roman2021dielectric}%
  \BibitemOpen
  \bibfield  {author} {\bibinfo {author} {\bibfnamefont {L.}~\bibnamefont
  {Rom\'an~Castellanos}}, \bibinfo {author} {\bibfnamefont {O.}~\bibnamefont
  {Hess}}, \ and\ \bibinfo {author} {\bibfnamefont {J.}~\bibnamefont
  {Lischner}},\ }\href@noop {} {\bibfield  {journal} {\bibinfo  {journal} {The
  Journal of Physical Chemistry C}\ }\textbf {\bibinfo {volume} {125}},\
  \bibinfo {pages} {3081} (\bibinfo {year} {2021})}\BibitemShut {NoStop}%
\bibitem [{\citenamefont {Otsuki}\ \emph {et~al.}(2021)\citenamefont {Otsuki},
  \citenamefont {Sugawa},\ and\ \citenamefont {Jin}}]{Otsuki2021}%
  \BibitemOpen
  \bibfield  {author} {\bibinfo {author} {\bibfnamefont {J.}~\bibnamefont
  {Otsuki}}, \bibinfo {author} {\bibfnamefont {K.}~\bibnamefont {Sugawa}}, \
  and\ \bibinfo {author} {\bibfnamefont {S.}~\bibnamefont {Jin}},\ }\href@noop
  {} {\bibfield  {journal} {\bibinfo  {journal} {Materials Advances}\ }\textbf
  {\bibinfo {volume} {2}},\ \bibinfo {pages} {32} (\bibinfo {year}
  {2021})}\BibitemShut {NoStop}%
\bibitem [{\citenamefont {Haynes}(2015)}]{haynes2015crc}%
  \BibitemOpen
  \bibfield  {author} {\bibinfo {author} {\bibfnamefont {W.}~\bibnamefont
  {Haynes}},\ }\href@noop {} {\emph {\bibinfo {title} {CRC handbook of
  chemistry and physics : a ready-reference book of chemical and physical
  data}}}\ (\bibinfo  {publisher} {CRC Press},\ \bibinfo {address} {Boca Raton,
  Florida},\ \bibinfo {year} {2015})\BibitemShut {NoStop}%
\bibitem [{\citenamefont {Blum}\ \emph {et~al.}(2009)\citenamefont {Blum},
  \citenamefont {Gehrke}, \citenamefont {Hanke}, \citenamefont {Havu},
  \citenamefont {Havu}, \citenamefont {Ren}, \citenamefont {Reuter},\ and\
  \citenamefont {Scheffler}}]{Blum2009}%
  \BibitemOpen
  \bibfield  {author} {\bibinfo {author} {\bibfnamefont {V.}~\bibnamefont
  {Blum}}, \bibinfo {author} {\bibfnamefont {R.}~\bibnamefont {Gehrke}},
  \bibinfo {author} {\bibfnamefont {F.}~\bibnamefont {Hanke}}, \bibinfo
  {author} {\bibfnamefont {P.}~\bibnamefont {Havu}}, \bibinfo {author}
  {\bibfnamefont {V.}~\bibnamefont {Havu}}, \bibinfo {author} {\bibfnamefont
  {X.}~\bibnamefont {Ren}}, \bibinfo {author} {\bibfnamefont {K.}~\bibnamefont
  {Reuter}}, \ and\ \bibinfo {author} {\bibfnamefont {M.}~\bibnamefont
  {Scheffler}},\ }\href@noop {} {\bibfield  {journal} {\bibinfo  {journal}
  {Computer Physics Communications}\ }\textbf {\bibinfo {volume} {180}},\
  \bibinfo {pages} {2175} (\bibinfo {year} {2009})}\BibitemShut {NoStop}%
\bibitem [{\citenamefont {Perdew}\ \emph {et~al.}(1996)\citenamefont {Perdew},
  \citenamefont {Burke},\ and\ \citenamefont {Ernzerhof}}]{Perdew1996}%
  \BibitemOpen
  \bibfield  {author} {\bibinfo {author} {\bibfnamefont {J.~P.}\ \bibnamefont
  {Perdew}}, \bibinfo {author} {\bibfnamefont {K.}~\bibnamefont {Burke}}, \
  and\ \bibinfo {author} {\bibfnamefont {M.}~\bibnamefont {Ernzerhof}},\
  }\href@noop {} {\bibfield  {journal} {\bibinfo  {journal} {Physical Review
  Letters}\ }\textbf {\bibinfo {volume} {77}},\ \bibinfo {pages} {3865}
  (\bibinfo {year} {1996})}\BibitemShut {NoStop}%
\bibitem [{\citenamefont {van Lenthe}\ \emph {et~al.}(1994)\citenamefont {van
  Lenthe}, \citenamefont {Baerends},\ and\ \citenamefont
  {Snijders}}]{vanLenthe1994}%
  \BibitemOpen
  \bibfield  {author} {\bibinfo {author} {\bibfnamefont {E.}~\bibnamefont {van
  Lenthe}}, \bibinfo {author} {\bibfnamefont {E.~J.}\ \bibnamefont {Baerends}},
  \ and\ \bibinfo {author} {\bibfnamefont {J.~G.}\ \bibnamefont {Snijders}},\
  }\href@noop {} {\bibfield  {journal} {\bibinfo  {journal} {The Journal of
  Chemical Physics}\ }\textbf {\bibinfo {volume} {101}},\ \bibinfo {pages}
  {9783} (\bibinfo {year} {1994})}\BibitemShut {NoStop}%
\end{thebibliography}

%
\end{document}